%% file: ifacconf.tex
\documentclass{ifacconf}

\usepackage{graphicx}      
\usepackage{natbib}        
\usepackage{tabularx}
\usepackage{amsmath}
\usepackage{enumitem}
\usepackage{booktabs}
\usepackage{multirow}
\usepackage{multicol}   
\usepackage{array}      
\usepackage{siunitx}
\usepackage[table]{xcolor}
\usepackage{url}

\begin{document}
\begin{frontmatter}

\title{A Multi-Worker Assembly Line Rebalancing with Spatial and Ergonomic Considerations\thanksref{footnoteinfo}} 

\thanks[footnoteinfo]{This work was supported by the EUREKA ITEA4 ArtWork project (2023-00970), and the Wallenberg AI, Autonomous Systems and Software Program (WASP) funded by the Knut and Alice Wallenberg Foundation.}

\author[First]{Martina Vinetti} 
\author[First]{Sabino F. Roselli} 
\author[First]{Martin Fabian}

\address[First]{Chalmers University of Technology, 
   Gothenburg, Sweden \\(e-mail: \{vinetti, rsabino, fabian\}@chalmers.se).}

\begin{abstract}                
This work addresses the Assembly Line Rebalancing Problem driven by cycle-time changes in manual assembly systems where multiple workers operate in parallel within the same station. A multi-objective optimization model is proposed that incorporates task reassignment, worker allocation, ergonomic evaluation, and explicit spatial feasibility through work-area constraints. The formulation minimizes deviations from the current configuration while promoting balanced workload and ergonomic conditions among workers. The main contribution is the extension of assembly line rebalancing to multi-worker settings with explicit spatial constraints. Computational experiments on synthetic instances demonstrate that the model consistently generates feasible reconfigurations, highlighting its potential as a decision-support tool for industrial rebalancing in flexible production environments.
\end{abstract}


\end{frontmatter}

\section{Introduction}
\input{Introd_NEW}

\section{Use Case}
\input{Usecase}

\section{Model Description}
\input{Math}

\section{Model Evaluation}
\input{Evaluation}

\section{Conclusion}
\input{Conclusion}

\bibliography{ifacconf}             
                                                   







\end{document}

%% file: Introd_NEW.tex
One of the pillars of \emph{Industry 4.0} is the shift from \emph{make-to-stock} to \emph{make-to-order} production systems~\citep{Ind4}, which enables higher customization but also increases variability in production volumes and product types. As a result, many manufacturers
face \emph{seasonal fluctuations in demand} occurring a few times throughout the year, which alter the required production rhythm and
may necessitate adjustments to the assembly line.

Assembly lines are typically organized as sequences of stations, each responsible for a subset of tasks. The \emph{cycle time} specifies the production rhythm by defining the interval at which each product advances to the next
station. Determining how tasks should be distributed across stations under a
given cycle time is the subject of the classical \emph{Assembly Line Balancing
Problem} (ALBP)~\citep{ALBP_1}. The ALBP seeks to assign tasks to stations while satisfying precedence constraints to optimize one or more performance criteria. The ALBP has been extensively studied, and numerous variants have been formulated and optimized using exact and heuristic methods~\citep{Becker2006,Boysen2022}.


In industrial practice, however, balancing an assembly line is not a one-time activity. Once a line has been initially designed and balanced, production changes, such as shifts in demand, product mix, technology, or organizational structure, require updating the allocation of tasks across stations. Redesigning the line from scratch each time is neither necessary nor economical. This challenge was first examined in early studies on reconfiguring existing assembly lines~\citep{ALRBP1} and later became widely known as the Assembly Line Rebalancing Problem (ALRBP). As pointed out by several authors~\citep{MSF1,MSF3}, significant modifications to the current configuration may require operator retraining, relocation of equipment, adjustments to material logistics, and changes to the workstation layout. Therefore, large deviations from the current configuration are generally discouraged, making it essential to evaluate not only the performance of the new solution but also its distance from the existing one. From an operational perspective, rebalancing is far more common than installing a new line, since changes in production requirements occur much more frequently than complete system redesigns. Despite this industrial relevance, and as highlighted in the survey by~\cite{ALRBP_Review_2022}, the ALRBP has received considerably less attention in the literature than the classical ALBP, and relatively few studies have addressed it in realistic industrial settings. ALRBP inherits the combinatorial structure of ALBP while incorporating additional constraints from the current configuration, it follows that ALRBP is likewise NP-hard.

To the best of our knowledge, \emph{assembly line rebalancing for multi-worker assembly lines} has only recently begun to be considered in the literature~\citep{Camli2025}. This arrangement is typical in large-vehicle assembly, where the size and complexity of the product require several operators in each station. These systems—referred to in the literature as multi-manned or multi-worker stations—introduce additional coordination challenges not present in single-worker stations. In multi-worker settings, rebalancing cannot be limited to reallocating tasks among stations; it must also include the assignment of tasks to individual workers
(e.g.,~Roshani and Giglio, 2017). Indeed, the effective processing capability of a station depends on the number of operators assigned to it, and therefore the task--worker assignment becomes a necessary decision dimension in rebalancing multi-worker stations.

Moreover, in flexible production environments, assembly work is predominantly performed by humans. Therefore, \emph{ergonomic aspects} must be carefully addressed, as repetitive and physically demanding tasks can cause long-term musculoskeletal disorders, as reported in~\cite{ErgRISKS}. Over the past decades, ergonomics has been increasingly integrated into ALBP~\citep{ErgALBP} and ALRBP~\citep{ErgALRBP} to ensure a fair distribution of physical effort among operators and support productivity and long-term sustainability. Consequently, task--worker assignment becomes essential to ensure balanced ergonomic effort across workers.


A further critical aspect in multi-worker stations is the presence of \emph{distinct working areas} within each station. Because multiple workers operate in parallel on the same product, spatial compatibility determines whether tasks can be performed
simultaneously without interference. Ignoring such work-area constraints
may yield allocations that are infeasible or unsafe, even when cycle time and
precedence relations are satisfied. The relevance of spatial considerations in
multi-worker assembly has been emphasized in recent studies
(e.g.,~\cite{Pilati,Yang}), highlighting the need to model working areas explicitly when rebalancing multi-worker stations.

In this work, a \emph{multi-objective optimization-based framework} for assembly line rebalancing in multi-worker systems is proposed. The approach integrates task allocation, worker assignment, and ergonomic aspects, producing feasible reconfigurations that minimize changes to the current configuration while ensuring an equitable distribution of workload and ergonomic effort among workers. This enables human-centered rebalancing decisions in dynamic manufacturing environments where adaptability, efficiency, and worker well-being must coexist.

%% file: Usecase.tex
The proposed framework has been developed in collaboration with a construction-equipment manufacturer to automate the rebalancing of one of its assembly lines. The line is currently rebalanced manually to address seasonal demand changes, relying on the expertise of production engineers to redistribute tasks and workers across stations. Although this expertise-based approach has proven effective in maintaining operational continuity, it is time-consuming, lacks standardization, and typically explore only a limited set of reallocation options—often resulting in suboptimal line performance.

Despite these limitations, manual rebalancing remains the standard practice. This is partly due to the effort required to maintain the detailed data needed for optimization, and partly because existing automation tools do not fully capture realistic industrial requirements such as multi-worker stations, spatial feasibility, and ergonomic considerations.

The industrial use case is the hood-assembly line of the construction-equipment plant. Workstations are organized as multi-worker units, where workers operate in parallel on the same workpiece while respecting spatial separation and using dedicated tools. This configuration enables high flexibility in task execution but substantially increases the complexity of rebalancing decisions. As a result, the line offers an industrially relevant environment for evaluating advanced, human-centered rebalancing methods.


%% file: Math.tex
Consider an assembly line consisting of a set of stations $S$, staffed by workers in the set 
$W$, and responsible for executing a set of indivisible tasks $T$. The line was originally balanced under a previous production plan, but due to variations in product demand, the required cycle time $CT$ has changed. To maintain efficiency, the line must be rebalanced while satisfying precedence and spatial constraints.

Each workstation may host multiple workers operating in parallel, each confined to a dedicated work-area 
(internal or external to the workpiece)
without exchanging tools or switching areas. Each task $i$ is characterized by a known processing time $\tau_i$, an ergonomic index
$e_i \in \{1,\dots,5\}$, and a work-area attribute $a_i \in \{0,1\}$, where $a_i = 0$ denotes external tasks and $a_i = 1$ denotes internal tasks.
Higher values of $e_i$ correspond to greater physical load, with $e_i=1$ indicating low ergonomic risk and $e_i=5$ indicating high ergonomic risk. Precedence constraints specify the ordering among tasks, ensuring that required tasks are completed before their successors can begin. 
These elements define the structure of the current line configuration, which is used
as a reference to minimize deviation in the rebalancing process.

The objective of the proposed model is to determine a new feasible configuration that satisfies the new cycle time, precedence, and work-area constraints while minimizing  
(i) deviations from the current configuration, 
(ii) workload imbalance across workers, and  
(iii) ergonomic imbalance across workers, the latter two of which indirectly affect station-level balance. 

The problem setting is based on the following assumptions:
\begin{enumerate}[label=\alph*)]
    \item Each task can be assigned to any station but must be executed in exactly one station and by one worker.
    \item Each workstation corresponds to the assembly of a single workpiece (e.g., a hood) and includes two distinct work areas (internal and external).
    \item Workers in the same station act in parallel and independently, without exchanging tools or switching work-areas.
    \item Each station may host multiple workers, and all stations are assumed to offer equivalent working space. Since the model does not incorporate physical layout constraints or task–station compatibility restrictions that would naturally limit staffing, lower and upper bounds on the number of workers per station are defined as $\underline{w}=\big\lfloor |W|/|S| \big\rfloor$ and $\overline{w}=\big\lceil |W|/|S| \big\rceil$, respectively.


\end{enumerate}

The notation used in the formulation of the problem is summarized below.

\noindent\textbf{Sets}

\begin{tabularx}{\linewidth}{lX}
$T$ & set of tasks, denoted by $i$ and $j$.\\
$S$ & set of stations, denoted by $s$.\\
$W$ & set of workers, denoted by $w$.\\
$\pi_i$ & set of immediate predecessors of task $i$.\\
$N_i$ & set of tasks sharing the same station as task $i$ in the current configuration. \\
\end{tabularx}

\noindent\textbf{Decision Variables}
\vspace{-0.2cm}

\medskip
\noindent\emph{Binary variables}

\begin{tabularx}{\linewidth}{lX}
$x_{iw}$   & 1 if task $i$ is assigned to worker $w$, 0 otherwise.\\
$y_{sw}$   & 1 if worker $w$ is assigned to station $s$, 0 otherwise.\\
$z_{is}$   & 1 if task $i$ is assigned to station $s$, 0 otherwise.\\
$s_s$      & 1 if station $s$ hosts multiple workers, 0 otherwise.\\
$u_w$      & 1 if worker $w$ operates in a shared station, 0 otherwise.\\
$q_{ijs}$  & 1 if tasks $i$ and $j$ are co-assigned at station $s$.\\
\end{tabularx}

\medskip
\noindent\emph{Auxiliary continuous variables}

\begin{tabularx}{\linewidth}{lX}
$l_w$     & Workload of worker $w$.\\
$h_w$     & Ergonomic load of worker $w$.\\
$l_{\max}, l_{\min}$ & Maximum and minimum workloads.\\
$h_{\max}, h_{\min}$ & Maximum and minimum ergonomic loads.\\
\end{tabularx}

Using these variables, the rebalancing problem is formulated as a multi-objective optimization model that simultaneously addresses configuration similarity, workload balance, and ergonomic balance.

\subsection{Mathematical Formulation}

\newcommand{\SF}{\ensuremath{\mathit{SF}}}

To quantify the deviation between the current and the new configuration, the model uses the \textit{Mean Similarity Factor} (MSF) proposed by \cite{MSF1}. This metric quantifies the preservation of the current configuration by
measuring, for each task $i$, the proportion of tasks that were previously assigned to the same station and remain together after rebalancing. The MSF is defined as:


\begin{equation}
\mathrm{MSF} = \frac{1}{|T|} \sum_{i=1}^{|T|} \SF_i.
\label{eq:MSF}
\end{equation}

\newcommand{\TIB}{\ensuremath{\mathit{TIB}}}
\newcommand{\TNB}{\ensuremath{\mathit{TNB}}}

Let $\TIB$ and $\TNB$ denote the sets of tasks, excluding $i$, assigned to the same station as task $i$ in the current and new configuration, respectively. 
The similarity factor for task $i$ is then expressed as:

\begin{equation}
\SF_i = \frac{|\;\TIB_i \cap \TNB_i\;|}{|\;\TIB_i\;|}.
\label{eq:SFi}
\end{equation}

The MSF ranges between 0 and 1, where values close to~1 indicate a high similarity between the two configurations, while lower values correspond to larger deviations. 
Hence, maximizing MSF minimizes the number of task reassignments during rebalancing, preserving the continuity of the current configuration while adapting the line to new production requirements.

In addition to preserving similarity with the current configuration, the model promotes a balanced allocation of both workload and ergonomic load across workers. For each worker $w$, the total workload and ergonomic load are defined as:
\begin{align}
l_w &= \sum_{i \in T} \tau_i \, x_{iw}, \label{eq:workload}\\
h_w &= \sum_{i \in T} e_i \, x_{iw}. \label{eq:ergonomic}
\end{align}

To quantify imbalance, auxiliary variables $l_{\max}$, $l_{\min}$, $h_{\max}$, and $h_{\min}$ bound
the workload and ergonomic load 
for each worker:
\begin{align}
l_{\min} \le l_w \le l_{\max}, \qquad
h_{\min} \le h_w \le h_{\max}.
\end{align}

The corresponding dispersion measures are given by:
\begin{align}
\Delta l &= l_{\max} - l_{\min}, \label{eq:workload_diff}\\
\Delta h &= h_{\max} - h_{\min}. \label{eq:ergonomic_diff}
\end{align}

Minimizing these two ranges promotes a uniform utilization of human resources and a fair distribution of physical effort, contributing to both operational efficiency and worker well-being.

Accordingly, the rebalancing problem is formulated as a multi-objective optimization model:
\begin{align}
\min \;\Big( -\mathrm{MSF} \;+\; \Delta l \;+\; \Delta h \Big).
\label{eq:obj}
\end{align}

The model is subject to a set of constraints that ensure consistency among assignments, compliance with precedence relations, and feasibility with respect to temporal and spatial limitations. 
The complete formulation is given below:
\allowdisplaybreaks
\begin{flalign}
&\sum_{w\in W} x_{iw} = 1 &&& \forall i\in T \label{eq:task-worker}\\
&\sum_{s\in S} z_{is} = 1 &&& \forall i\in T \label{eq:task-station}\\
&\sum_{s\in S} y_{sw} \leq 1 &&& \forall w\in W \label{eq:worker-once}\\
&x_{iw} + y_{sw} \leq  1 + z_{is} &&& \forall i\in T,\, s\in S,\, w\in W \label{eq:consistency}\\
&\underline{w} \leq \sum_{w\in W} y_{sw} \leq \overline{w} &&& \forall s\in S \label{eq:station-balance}\\
&z_{is_1} + z_{js_2} \leq 1 \phantom{\sum_{w\in W} y_{sw}} &&& \forall j\in T,\; i\in \pi_{j},\; s_1>s_2 \label{eq:precedence}\\
&l_w \leq \mathrm{CT} \phantom{\sum_{w\in W} y_{sw}} &&& \forall w\in W \label{eq:cycle-time}\\
&\sum_{w\in W} y_{sw} \geq 2 s_s &&& \forall s\in S \label{eq:shared-lb}
\end{flalign}
\begin{flalign}
&\sum_{w\in W} y_{sw} \leq 1 + (|W|-1) s_s &&& \forall s\in S \label{eq:shared-ub}\\
&u_w \geq \tfrac{1}{|S|}\sum_{s\in S} s_s y_{sw} &&& \forall w\in W \label{eq:uw-def}
\end{flalign}
\begin{flalign}
&\sum_{i\in T} a_i x_{iw}
 \sum_{i\in T} (1-a_i)x_{iw} 
 \leq (1-u_w)|T|^2  &&  \forall w\in W \label{eq:mismatch}
\end{flalign}
\begin{flalign}
&q_{ijs} = z_{is} z_{js}  &&& \forall i\in T,\, j\in N_i,\, s\in S \label{eq:coassign}
\end{flalign}

Constraints~(\ref{eq:task-worker})--(\ref{eq:task-station}) enforce assignment exclusivity; each task is assigned to exactly one worker and exactly one station.
Constraint~(\ref{eq:worker-once}) restricts each worker to at most one station, while~(\ref{eq:consistency}) 
guarantees that if worker $w$ does task $i$ in station $s$, then task $i$ is assigned to station $s$.
Constraint~(\ref{eq:station-balance}) controls the number of workers per station, keeping it within the specified bounds $[\underline{w}, \overline{w}]$. 
Task precedence is enforced by~(\ref{eq:precedence}), preventing any successor from being assigned to an earlier station than its predecessor.

Constraint~(\ref{eq:cycle-time}) ensures that the total workload of each worker does not exceed the cycle time $\mathrm{CT}$. 

Constraints~(\ref{eq:shared-lb})--(\ref{eq:shared-ub}) define $s_s$ as $1$ when station $s$ hosts multiple workers and $0$ otherwise. Constraint~(\ref{eq:uw-def}) assigns $u_w = 1$ to workers operating in shared stations.

Constraint~(\ref{eq:mismatch}) enforces the work-area rule by restricting
task--worker assignments based on the task’s area attribute
$a_i\in\{0,1\}$ (internal if $a_i=1$, external if $a_i=0$),
through a quadratic term that prevents a worker in a shared station from
executing tasks belonging to different areas.
The bilinear term becomes positive only if worker $w$ is assigned tasks from both areas. The constraint forbids this whenever $u_w=1$ (shared station), thereby
ensuring area consistency. When $u_w=0$, the constraint is non-binding since a
single worker in a station may execute tasks from both areas.


Finally, constraint~(\ref{eq:coassign}) defines the co-assignment variable $q_{ijs}$,
which equals $1$ when tasks $i$ and $j$, assigned to the same station
in the current configuration, remain together after rebalancing. The variables $q_{ijs}$ directly contribute to the computation of the MSF.

Constraints~(\ref{eq:mismatch}) and~(\ref{eq:coassign}) include bilinear binary
terms and are therefore linearized to improve tractability.

\subsection{Linearization of Nonlinear Constraints}

Bilinear relations in constraints~(\ref{eq:mismatch}) and~(\ref{eq:coassign}) are replaced with equivalent MILP-compatible linear encodings, ensuring the same set of integer-feasible assignments.

Since constraint~(\ref{eq:coassign}) involves the bilinear term $z_{is}z_{js}$, the
co-assignment variable $q_{ijs}$ is enforced through the standard linearization:
\begin{align}
q_{ijs} &\le z_{is}, \tag{\ref{eq:coassign}a} \label{eq:coassign-a}\\
q_{ijs} &\le z_{js}, \tag{\ref{eq:coassign}b} \label{eq:coassign-b}\\
q_{ijs} &\ge z_{is} + z_{js} - 1. \tag{\ref{eq:coassign}c} \label{eq:coassign-c}
\end{align}

To linearize~(\ref{eq:mismatch}), two new auxiliary binaries are introduced:
$c_w$ specifies the working area assigned to worker $w$ ($1$ = internal,
$0$ = external), and $\ell_{sw}$ equals $1$ if $w$ is the sole worker at station
$s$.
The work-area logic is then encoded through the following per-task inequalities:
\begin{flalign}
&x_{iw} \le c_w + (1 - y_{sw}) + (1 - z_{is}) + \ell_{sw}, &&\nonumber 
\end{flalign}
\vspace{-0.52cm}
\begin{flalign}
&&\forall i \in T : a_i = 1,\; s \in S,\; w \in W,
\label{eq:old-mismatch1}
\end{flalign}
\begin{flalign}
&x_{iw} \le (1 - c_w) + (1 - y_{sw}) + (1 - z_{is}) + \ell_{sw},  && \nonumber  
\end{flalign}
\vspace{-0.52cm}
\begin{flalign}
&& \forall i \in T : a_i = 0,\; s \in S,\; w \in W.
\label{eq:old-mismatch2}
\end{flalign}
When worker $w$ is assigned to station $s$ and executes task $i$ (i.e., $x_{iw}=1$,
$y_{sw}=1$, $z_{is}=1$), and the station is shared ($\ell_{sw}=0$), these
constraints force $a_w$ to match the area of task $i$, thereby preventing
a worker in a shared station from simultaneously handling internal and external
tasks. If $\ell_{sw}=1$, meaning that $w$ is alone at station $s$, the inequalities
are non-restrictive and mixed-area assignments are allowed.

This linearized formulation can be shown to be equivalent to the quadratic
one, in the sense that both formulations yield the same possible assignments for the decision variables $(x,y,z)$. A constructive equivalence proof can be obtained by  
(i) deriving $(s_s,u_w)$ from any feasible solution of the linearized model and
verifying~(\ref{eq:mismatch}); and  
(ii) conversely, deriving $(c_w,\ell_{sw})$ from any feasible solution of the
quadratic model and verifying~(\ref{eq:old-mismatch1})--(\ref{eq:old-mismatch2}).

This equivalence concerns exclusively the set of integer feasible assignments 
in the $(x,y,z)$ space. The two encodings need not yield identical LP relaxations, 
as will be demonstrated in the computational analysis.

%% file: Evaluation.tex
The proposed model introduces a work-area constraint that, to the authors' knowledge, has not previously been incorporated into the ALRBP. Since both a quadratic and a linear formulation of this constraint were developed, the evaluation aimed to assess their relative computational performance. Preliminary tests indicated that the work-area constraint is the principal driver of solution difficulty, since solution times drop when this constraint is relaxed. This observation further motivates a systematic comparison of the two formulations.


To assess the model performance across a range of problem sizes and structural characteristics, a set of synthetic instances was generated. Using established ALBP benchmarks was deemed inappropriate, as these datasets do not include work-area information and would therefore not allow meaningful computational or qualitative comparisons. Synthetic instances were generated using the cycle time of the industrial hood-line (20 minutes) as reference, with problem sizes ranging from 10 to 40 tasks. For each task,  processing times were sampled uniformly from $[1,7]$ minutes, ergonomic scores from $\{1,\dots,5\}$, and working areas were assigned at random. Precedence relations were also generated randomly while ensuring feasibility; for each task $j$, up to three predecessors were selected among tasks $\{1,\dots,j-1\}$, which guarantees that all precedence relations remain acyclic.

Since generating a random current configuration would not guarantee feasibility with respect to cycle time, work-area and precedence constraints, each synthetic instance was first solved under a cycle time different from the one used for rebalancing (i.e., any value in the range $[17,23]$ minutes excluding $20$). In this preliminary optimization, only the workload- and ergonomic-balancing objectives were considered. The result provides a feasible current configuration for the subsequent rebalancing at the target cycle time of $20$ minutes.

For both the initial balancing and the rebalancing, the number of workers was set to the minimum value ensuring feasibility. This choice forces workloads to be as close as possible to the cycle time, reflecting realistic production conditions. As a consequence, the number of workers may differ between the current and rebalanced configurations.


Since all three objectives are considered equally relevant for the rebalancing problem, identical weights of $1/3$ were assigned in the experiments, providing a neutral baseline in the absence of a priori preferences. A preliminary sensitivity analysis showed that varying the weights often leads to identical optimal solutions, due to the constrained structure of the feasible solution space. However, the objectives exhibit different scales; the MSF is naturally bounded by $[0,1]$, whereas the workload range can vary up to $\text{CT}-1$ (because, under the given formulation, each worker is assigned at least one task), and the ergonomic range is potentially unbounded from above. To allow a meaningful aggregation, all objectives were normalized to the interval $[0,1]$. Normalization was performed using the Nadir–Utopia method \citep{normalization}, in which each objective value $f_k$ is
transformed according to:
\[
f_k^{\mathrm{norm}} = 
\frac{f_k - f_k^{\mathrm{Utopia}}}{f_k^{\mathrm{Nadir}} - f_k^{\mathrm{Utopia}}}.
\]
This ensures commensurability among the three criteria and prevents any single objective from dominating the weighted sum due to scale differences.

The Nadir and Utopia values for each objective were obtained empirically by solving the model three times, each time optimizing a single objective while keeping all constraints active. For each criterion, the best and worst values observed
across the three runs were used as the Utopia and Nadir points, respectively,
and applied for objective normalization in all experiments.

Two alternative formulations were implemented depending on how the
work-area constraint is modeled, a quadratic version, which yields a
mixed-integer quadratic program (MIQP), and a linearized version, resulting
in a mixed-integer linear program (MILP). Both were solved using exact
optimization to establish a reliable baseline of their computational behavior. This choice reflects the exploratory nature of the present study; the goal is to evaluate the intrinsic difficulty of the proposed formulations rather than to develop a scalable solution method.

All experiments were conducted using the commercial solver \emph{Gurobi}
(version~12.0.2), which is widely used for exact optimization in MILP and MIQP
settings and is recognized for its state-of-the-art branch-and-bound and
cutting-plane algorithms \citep{gurobi}. Although Gurobi is not open-source, it
remains one of the most established and consistently benchmarked solvers for
mixed-integer optimization, making it suitable for a first computational
evaluation of the proposed model.

Experiments were executed on an Apple Mac Studio~(2025) with an Apple M4 Max processor and 64\,GB RAM, under macOS~Tahoe~26.1. A three-hour time limit was applied to all runs as a practical upper bound for experimental evaluation%
\footnote{Code and experimental problem instances are available at: \url{https://github.com/Chalmers-Control-Automation-Mechatronics/ALRP}}.

\begin{figure}[h]
    \centering
    \includegraphics[width=0.45\textwidth]{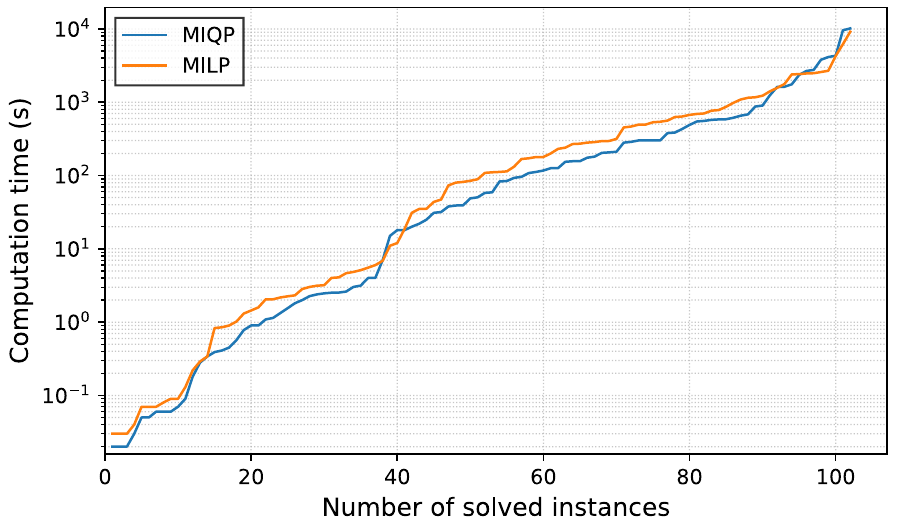}
    \caption{Comparison of computation times between the MIQP and MILP formulations.}
    \label{fig:caPlot}
\end{figure}

Figure~\ref{fig:caPlot} shows the computational comparison of the two formulations across all instances solved within the time limit.
The benchmark was initially composed of 70 synthetic instances, generated for problem sizes of 10, 15, 20, 25, 30, 35, and 40 tasks, with ten instances for each size. Preliminary experiments on this set revealed a clear shift in the relative performance of the two formulations when moving from 30 to 35 tasks. To better analyze this transition, an additional set of 40 instances with 31–34 tasks was generated to refine the benchmark around the region where the change in trend occurred.

In Figure~\ref{fig:caPlot}, the instances are ordered by increasing solution time, and the horizontal axis reflects the number of instances successfully solved within the time limit.  The two approaches deliver broadly similar performance: MIQP achieves faster solution times on a larger subset of the benchmark, whereas MILP becomes more efficient on many of the most challenging instances that remain solvable within the time limit. Across the entire benchmark, computation times also exhibit substantial variability, even among instances of the same size, reflecting differences in the underlying structural complexity of individual problem instances.

A complementary aspect of the evaluation is the quality of the rebalanced configurations. Table~\ref{tab:fairness_summary} summarizes the behaviour of the three optimization objectives, MSF, workload fairness (WL), and ergonomic fairness (EL), across increasing problem sizes (10–40 tasks). Fairness is quantified using two dispersion measures, the Normalized Range (NR), which compares the difference between the most and least loaded worker to the average load, and the Coefficient of Variation (CV), which relates the standard deviation to the mean. Both NR and CV can be interpreted as percentage deviations from the average load, with lower values indicating more equitable distributions.

\begin{table}[t]
\centering
\renewcommand{\arraystretch}{1.2}
\setlength{\tabcolsep}{6pt}
\caption{Similarity and fairness metrics across problem sizes.}
\begin{tabular}{l c cc cc}
\toprule
\multirow{2}{*}{\textbf{Problem Size}} 
 & \multirow{2}{*}{\textbf{MSF}} 
 & \multicolumn{2}{c}{\textbf{WL}} 
 & \multicolumn{2}{c}{\textbf{EL}} \\
\cmidrule(lr){3-4} \cmidrule(lr){5-6}
 & & \textbf{NR} & \textbf{CV} & \textbf{NR} & \textbf{CV} \\
\midrule
10 & 0.788 & 0.427 & 0.102 & 0.385 & 0.088 \\
20 & 0.631 & 0.311 & 0.089 & 0.332 & 0.078 \\
30 & 0.461 & 0.272 & 0.069 & 0.331 & 0.070 \\
40 & 0.727 & 0.229 & 0.055 & 0.417 & 0.090 \\
\bottomrule
\end{tabular}
\label{tab:fairness_summary}
\end{table}

Overall, fairness improves as the number of tasks increases; both WL and EL show markedly lower NR and CV for instances with 30 and 40 tasks, suggesting that the additional degrees of freedom enable a more homogeneous allocation of work. In contrast, instances with only 10 tasks display the highest dispersion, due to limited combinatorial flexibility. The MSF does not follow a monotonic pattern with problem size. Medium-sized instances (20–30 tasks) require more substantial reallocations to achieve balanced workload and ergonomic conditions, leading to lower MSF values. Conversely, for 10-task instances the high MSF reflects the limited number of feasible reallocations, whereas 40-task instances benefit from increased solution-space flexibility, enabling fairness improvements while remaining close to the current task allocation.

To further assess the robustness of the approach, a dedicated analysis was performed for the instances with 30, 35, and 40 tasks, which exhibited the most informative fairness patterns. Two different current configurations are evaluated, an \emph{optimal start}, obtained by solving the initial balancing problem to optimality, and a \emph{suboptimal start}, generated by stopping the solver at an 80\% MIP gap. Table~\ref{tab:robustness} reports the corresponding similarity and fairness metrics before and after rebalancing.

\begin{table}[t]
\centering
\renewcommand{\arraystretch}{1.2}
\setlength{\tabcolsep}{6pt}
\caption{Fairness robustness across optimal and suboptimal incumbent solutions.}
\begin{tabular}{l c cc cc}
\toprule
\multirow{2}{*}{\textbf{Scenario}} 
 & \multirow{2}{*}{\textbf{MSF}} 
 & \multicolumn{2}{c}{\textbf{WL}} 
 & \multicolumn{2}{c}{\textbf{EL}} \\
\cmidrule(lr){3-4} \cmidrule(lr){5-6}
 & & \textbf{NR} & \textbf{CV} & \textbf{NR} & \textbf{CV} \\
\midrule
Optimal Start                 & --    & 0.088 & 0.038 & 0.137 & 0.055 \\
Rebalancing Opt     & 0.639 & 0.090 & 0.037 & 0.120 & 0.047 \\
Suboptimal Start              & --    & 0.170 & 0.063 & 0.360 & 0.126 \\
Rebalancing Subopt   & 0.494 & 0.094 & 0.038 & 0.123 & 0.049 \\
\bottomrule
\end{tabular}
\label{tab:robustness}
\end{table}


As expected, solutions derived from the optimal start show relatively low WL and EL dispersion, whereas the suboptimal start exhibits noticeably worse fairness values. After rebalancing, both scenarios achieve comparably balanced outcomes, with WL and EL metrics close to those obtained from the optimal start. MSF values behave as expected. Improving fairness from a suboptimal starting configuration requires larger changes, reflected by lower MSF, whereas starting from an optimal allocation preserves more of the current plan. Nevertheless, the resulting similarity levels indicate that substantial fairness improvements can be achieved without incurring excessive task reallocation.




%% file: Conclusion.tex
This study extends the 
assembly line rebalancing problem
by integrating two key industrial features, multi-worker stations and spatial feasibility through work-area constraints, which have only recently begun to be considered in the literature and have not been addressed jointly. The formulation also incorporates workload and ergonomic balance to promote long-term sustainability and worker well-being.

The results show that the proposed model provides rebalanced configurations that remain close to 
existing configurations,
while ensuring a fair distribution of workload and ergonomic effort among operators, regardless of the initial configuration.

Although computational performance currently limits scalability, the approach shows strong potential for industrial deployment, with future work focused on real-world validation and performance improvements, including the development of scalable solution approaches.